\documentclass[showpacs,preprintnumbers,twocolumn,amsmath,amssymb,prl]{revtex4}

\usepackage{graphicx}
\usepackage{dcolumn}
\usepackage{bm}

\begin{document}
\title{The influence of pairing on the nuclear matrix elements
       of the neutrinoless $\beta \beta$ decays }

\author{E. Caurier $^{*}$, J.~Men\'{e}ndez $^{+}$, F. Nowacki $^{*}$
and A.~Poves $^{+}$}

\affiliation{({*}) IPHC, IN2P3-CNRS/Universit\'{e} Louis Pasteur
BP 28, F-67037 Strasbourg Cedex 2, France\\
 (+) Departamento de F\'{\i}sica Te\'{o}rica,  Universidad Aut\'{o}noma
de Madrid and  Instituto de F\'{\i}sica Te\'{o}rica, UAM/CSIC, E-28049, Madrid, Spain}

\date{\today}

\begin{abstract}   
We study in this letter the neutrinoless double beta decay nuclear
matrix elements (NME's) in the framework of the Interacting Shell
Model. We analyze them in terms of the total angular momentum of the
decaying neutron pair and as a function of the seniority truncations
in the nuclear wave functions. This point of view turns out to be very
adequate to gauge the accuracy of the NME's predicted by different
nuclear models.  In addition, it gives back the protagonist role in
this process to the pairing interaction, the one which is responsible
for the very existence of double beta decay emitters.  We show that
low seniority approximations, comparable to those implicit in the
quasiparticle RPA in an spherical basis, tend to overestimate the
NME's in several decays.
  
\end{abstract}


\maketitle

The discovery of the massive character of the neutrinos 
in the recent measurement at 
Super-Kamiokande~\cite{Fukuda.Hayakawa.ea:1998b},
SNO~\cite{Ahmad.Allen.ea:2002a}
and KamLAND~\cite{Eguchi.Enomoto.ea:2003}, has openeded a new era in the
neutrino physics. However, these experiments are sensitive only
to the mass differences between the three neutrino species. Their
absolute mass scale and hierarchy are still unknown. In addition,
we don't know either if the neutrinos are Dirac or Majorana particles.
The double beta decay is the rarest nuclear weak process.  It takes
place between two even-even isobars, when the decay to the
intermediate nucleus is energetically forbidden due to the pairing
interaction, that shifts the even-even and the odd-odd mass parabolas
in a given isobaric chain.  The two-neutrino decay is just a second
order process in the weak interaction. It conserves the lepton number and
has been already observed in several nuclei.  A second mode, the
neutrinoless decay $0\nu \;\beta\beta$ can only take place if the neutrino
is a Majorana particle and demands an extension of the standard model of
electroweak interactions, because it violates the lepton number
conservation.  Therefore, the observation of the double beta decay
without emission of neutrinos will sign the Majorana character of the
neutrino and will establish the absolute mass scale of the neutrinos,
hence deciding their mass hierarchy.

The expression for the neutrinoless beta decay half-life, in the
$0^{+}\rightarrow 0^{+}$ case,  can be brought
to the following form 
\cite{Takasugi:1981,Doi.Kotani.Takasugi:85}:

\begin{equation} 
  [T_{1/2}^{(0\nu)}(0^{+}->0^{+})]^{-1}=
  G_{0\nu} \left( M^{(0\nu)} \left(\frac{\langle m_{\nu}\rangle}{m_{e}} \right) \right)^{2}
 \label{eq:0vh1}
\end{equation}

\begin{equation} 
  M^{(0\nu)} = M_{GT}^{(0\nu)} - \left(\frac{g_V}{g_A}\right)^{2}
   M_{F}^{(0\nu)} - M_{T}^{(0\nu)}
\label{eq:0vh2}
\end{equation}

\noindent
where $\langle m_{\nu} \rangle$ is the effective neutrino mass
(a linear combination of the neutrino mass eigenvalues
whose coefficients are the corresponding elements of the neutrino mixing matrix),
and $G_{0\nu}$ is the kinematic phase space factor \cite{suci}.

The important point at this stage is that, once the neutrinoless
double beta decay be detected, to transform the measured half life in
an accurate value of the effective neutrino mass would require a
precise computation of the nuclear matrix elements (NME's) of the
decay operators. This, in turn, demands a detailed description of the
structure of the nuclei involved in the process. A critical analysis
of the available predictions for the NME's of the potential
0$\nu$$\beta\beta$ emitters (only about one dozen) was made recently
by Bahcall {\it et al.} \cite{bahcall}. Their conclusion was rather
pessimistic, owing to the large dispersion of the calculated
values. In a subsequent paper, Rodin {\it et al.} \cite{rodin} have
shown that many of the quasi-particle RPA (QRPA) calculations taken
into account in Bahcall's survey were obsolete, and that, when these
are not considered, the spread of the calculated values is much
smaller. The aim of this work is to go one step further and to propose
a much narrower band of values for the NME's, based in the predictions
of large scale applications of the Interacting Shell Model (ISM) and
in the analysis of the QRPA results in terms of the pairing content of
their solutions.

The matrix elements M$_{GT,F,T}^{(0\nu)}$  can be
calculated in the closure approximation, that is good to better than
90\% due to the large average energy of the virtual neutrino
($\sim$100~MeV) \cite{sukhad}.  For the Gamow-Teller channel it reads,

\begin{equation}
 M_{GT}^{(0\nu)} = \langle 0_{f}^{+} | h(|\vec{r_1}-\vec{r_2}|) 
(\vec{\sigma}_1\cdot \vec{\sigma}_2) (t^-_1 \; t^-_2) | 0_{i}^{+} \rangle
\label{eqn:gt}
\end{equation}

\noindent
and similar expressions hold for the other matrix elements.  
$h(|\vec{r_1}-\vec{r_2}|)$ is the neutrino potential ($\sim$1/r)
obtained from the neutrino propagator. Higher order contributions (hoc)
to the nuclear current produce  the 
tensor term and add extra contributions to the 
Gamow-Teller expression of Eq.~(\ref{eqn:gt}) 
\cite{simko}.

Generically, the two body decay operators can be written in the Fock space representation as:

\begin{equation}
  \hat{M}^{(0\nu)} = \sum_J  \left( \sum_{i,j,k,l} M_{i,j,k,l}^J \; \left((a^{\dagger}_i a^{\dagger}_j)^J \;
                              (a_k a_l)^J \right)^0 \right),     
\end{equation}

\noindent
where the indices  {\it i, j, k, l,} run over the single particle orbits of the spherical nuclear mean field.
Applying the techniques of ref.~\cite{duzu} we can factorize the operators as follows

\begin{equation}
  \hat{M}^{(0\nu)} = \sum_{J^{\pi}}   \hat{P}^{\dagger}_{J^{\pi}} \; \;   \hat{P}_{J^{\pi}}     
\end{equation}

The operators $ \hat{P}_{J^{\pi}}$ annihilate pairs of neutrons coupled to $J^{\pi}$
in the father nucleus and the operators $ \hat{P}^{\dagger}_{J^{\pi}}$
substitute them by pairs of protons coupled to the same $J^{\pi}$. The overlap
of the resulting state with the ground state of the grand daughter
nucleus gives the $J^{\pi}$-contribution to the NME. The --a priori
complicated-- internal structure of these exchanged pairs is dictated
by the double beta decay operators.


 The ISM calculations reported in this letter are carried out in the
spirit of the previous shell model works
\cite{haxton,0nu1,0nu2,rmp}. For the A=76 and A=82 cases we make full
calculations in the valence space (1p$_{3/2}$, 0f$_{5/2}$, 1p$_{1/2}$,
0g$_{9/2}$) using a newly built effective interaction that, starting
with a G-matrix \cite{morten} has its matrix elements fitted to a
large set of experimental data. For the A=124, A=128, A=130, and A=136
emitters, we make full calculations in the valence space (0g$_{7/2}$,
1d$_{5/2}$, 1d$_{3/2}$, 2s$_{1/2}$, 0h$_{11/2}$) with another
interaction obtained in a similar manner.  These model spaces and
interactions will be discussed in detail elsewhere \cite{arnaud}. The
dimensions of the shell model bases reach in some cases
O(10$^{10}$). The present calculations adopt the closure
approximation, and model the short range and finite size corrections
as in \cite{0nu1}. We use r$_0$=1.2~fm to make the matrix elements
dimensionless and g$_A$=1.25.  The choice of g$_A$=1.25 instead of the
quenched value g$_A$=1.0 needed for the pure Gamow-Teller processes in
nuclei is consistent with the use of the closure approximation, in
which the multipole decomposition of the decay plays no role at all.
In addition, even without closure, the need of a quenched 
g$_A$ in the J$^{\pi}$=1$^+$ channel of the 0$\nu$ decay, that has no reason 
to resemble the pure Gamow-Teller operator of the 2$\nu$ decay, is not guaranteed.
  Higher order contributions to the nuclear current (hoc) \cite{simko}
are explicitly included for the first time in the ISM context, leading
to reductions of the NME's in the 15\% range.  Our final predictions
for $M^{(0\nu)}$ are gathered in table \ref{tab:update}.

\begin{table}
\begin{center}
\caption{ISM predictions for the 0$\nu$ double beta decay matrix elements, with and without higher 
 order contributions to the nuclear current (hoc). The effective neutrino mass corresponds to 
 T$_{\frac{1}{2}}$ = 10$^{25}$ y.\label{tab:update}}
    \begin{tabular*}{\linewidth}{@{\extracolsep{\fill}}cccc}
 \hline \\
  &  M$^{(0\nu)}$(no hoc) &  M$^{(0\nu)}$ & $\langle m_\nu\rangle$   \\ [5pt] \hline \\  
 $^{48}$Ca $\rightarrow$  $^{48}$Ti  & 0.76 & 0.59  & 1.09   \\
 $^{76}$Ge $\rightarrow$  $^{76}$Se  & 2.58 & 2.22  & 1.05   \\
 $^{82}$Se $\rightarrow$  $^{82}$Kr  & 2.49 & 2.11  & 0.50   \\
 $^{124}$Sn $\rightarrow$ $^{124}$Te & 2.38 & 2.02  & 0.53   \\
 $^{128}$Te $\rightarrow$ $^{128}$Xe & 2.67 & 2.26  & 2.27   \\
 $^{130}$Te $\rightarrow$ $^{130}$Xe & 2.41 & 2.04  & 0.41   \\
 $^{136}$Xe $\rightarrow$ $^{136}$Ba & 2.00 & 1.76  & 0.47   \\
\hline
\end{tabular*}
\end{center}
\end{table}

   Except for doubly magic $^{48}$Ca, whose NME  is severely
   quenched, our predictions cluster around a value
   M$^{(0\nu)}\approx$~2. The upper bounds on the neutrino mass for a half life
   of 10$^{25}$ y, that incorporate the phase space factors,
   show  a mild preference for  the potential emitters with A=82, 124, 130 and 136.
   The matrix elements are dominated by the Gamow-Teller
   contribution. 
 The influences of the restrictions in the valence space and of the
choice of the effective interaction in the ISM NME's have been studied
in \cite{nsa_epj}. In
all, these should result in a 20\% uncertainty of our predictions.
Treating the short range correlations with a prescription softer than
Jastrow might produce an increase of NME's, that we have not evaluated
yet in the ISM context, but we do not expect it to go beyond 10-15\%.

\begin{figure}[h]
 \begin{center}
    \includegraphics[width=1.0\columnwidth]{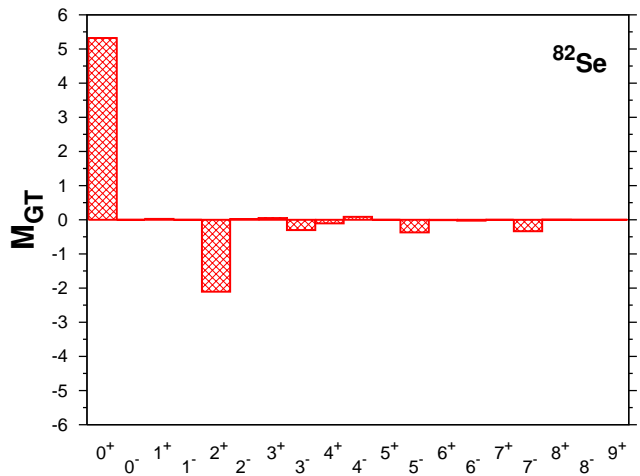}
\caption{(color online) Contributions to the Gamow Teller matrix element of the
          $^{82}$Se $\rightarrow$ $^{82}$Kr decay as a function of the J$^{\pi}$ of
          the transformed pair (no hoc).\label{fig:se82jj}}
  \end{center}
\end{figure}

   In order to explore the structure of the 0$\nu \beta \beta$ two
body transition operators, we have plotted in Figure \ref{fig:se82jj},
the contributions to the 0$\nu$ GT matrix element as a function of the
J$^{\pi}$ of the decaying pair.
The results are very suggestive, because the dominant contribution
corresponds to the decay of J=0 pairs, whereas the contributions of the
pairs with J$>$0 are either negligible or have opposite sign to the
leading one. 
This behavior is common to all the cases that we
have studied, as can be seen in table \ref{tab:pair_jj}. Notice 
that the cancellations are substantial.  These features are  also present
 in the QRPA calculations, in whose
context they had been discussed in
refs. \cite{engel,rodin}. 
\begin{table}[h]
\begin{center}
\caption{\label{tab:pair_jj}J=0 $vs$ J$>$0 pair contributions to the Gamow Teller matrix element (no hoc).}
    \begin{tabular*}{\linewidth}{@{\extracolsep{\fill}}llll}
\hline  \\
  &  M$_{GT}^{(0\nu)}$ &  M$_{GT}^{(0\nu)}$(J=0) & M$_{GT}^{(0\nu)}$(J$>$0)\\ [5pt] \hline\\ 
 $^{76}$Ge $\rightarrow$ $^{76}$Se & 2.35 &  5.59 & -3.24\\
 $^{82}$Se $\rightarrow$ $^{82}$Kr & 2.25 &  5.32 & -3.07 \\
 $^{130}$Te $\rightarrow$  $^{130}$Xe  & 2.12 &  6.58 & -4.46\\
 $^{136}$Xe $\rightarrow$ $^{136}$Ba & 1.77 &  5.72 & -3.95 \\
\hline 
\end{tabular*}
\end{center}
\end{table}

To grasp better this mechanism, we have expressed the matrix elements
in a basis of generalized seniority s (s counts the number of unpaired
nucleons in the nucleus); $ | 0_{i}^{+} \rangle = \sum_s \alpha_s | s
\rangle_i$ ; $ | 0_{f}^{+} \rangle = \sum_s \beta_s | s \rangle_f$.
The J=0 terms provide
essentially all the contribution to M$^{(0\nu)}$ that is diagonal in s.  The
canceling parts, J$>$0, produce almost exclusively cross terms with
$\Delta$s=+4. The matrix elements $ _f\langle s | \hat{M}^{(0\nu)} | s
\rangle_i $ are roughly proportional to (s$_{max} -$s), averaged in
parent and grand daughter, while the cross terms $ _f\langle s+4 |
\hat{M}^{(0\nu)} | s \rangle_i $ are constant --in both cases
scaled by the larger oscillator quantum number in each
valence space--.  The two body matrix elements of the operator 
$\hat{M}^{(0\nu)}(J=0)$ are almost identical to those of the isovector
pairing of the nuclear effective interaction, that is why it acts as a ``pair
counter''.  At present we cannot offer a similarly simple explanation
for the behavior of the J$>$0 terms.  Obviously, when the initial and
final states have seniority zero, the s=0 contribution is maximized
and the canceling terms are null, hence, M$^{(0\nu)}$ becomes 
maximal.

\begin{figure}[h] 
 \begin{center}
    \includegraphics[width=1.0\columnwidth]{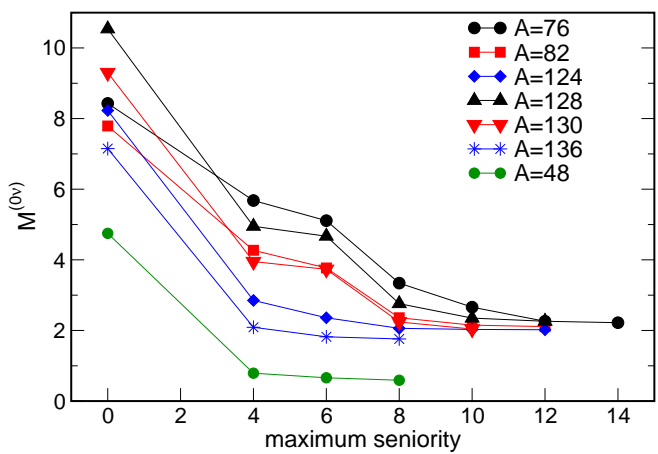}
\caption{\label{fig:s0nu}(color online)  The neutrinoless double beta decay NME's,  defined in equation
\ref{eq:0vh2}, as  a function of the
maximum seniority of the wave functions.}
  \end{center}
\end{figure}

These results highlight the role of the seniority structure of the
nuclear wave functions in the build-up of the 0$\nu$ NME's, and we
shall examine this issue for the competing theoretical
approaches.  In the first place, we have plotted the results of the 
ISM calculations of the NME's as a function of the seniority
in Fig.~\ref{fig:s0nu}. The values with maximum
seniority provide the exact ISM results in the corresponding valence
spaces. Two aspects are worth to underline; a) the strong reduction of
the NME as the maximum allowed seniority increases (up to a factor
five); and b) the fact that, at s$\le$4, the NME's of the  A=76,
82, 128, and 130 decays miss convergence by factors 2-3. On the
contrary, in the A=48, A=124, and 136 cases the convergence at s$\le$4
is much better.  The reason why these decays behave differently is
very illuminating; $^{124}$Sn has only neutrons in the valence space,
hence, its wave function is dominated by low seniority components and
its NME at s$\le$4 is quite close to the exact result; in the A=136
decay, the s$\le$4 calculation for $^{136}$Xe is exact, therefore, at
s$\le$4, the NME is also  close to the exact one; finally, in the A=48
decay the s$>$4 components are negligible both in doubly magic
$^{48}$Ca and in $^{48}$Ti.

\begin{figure}[h]
 \begin{center}
    \includegraphics[width=1.0\columnwidth]{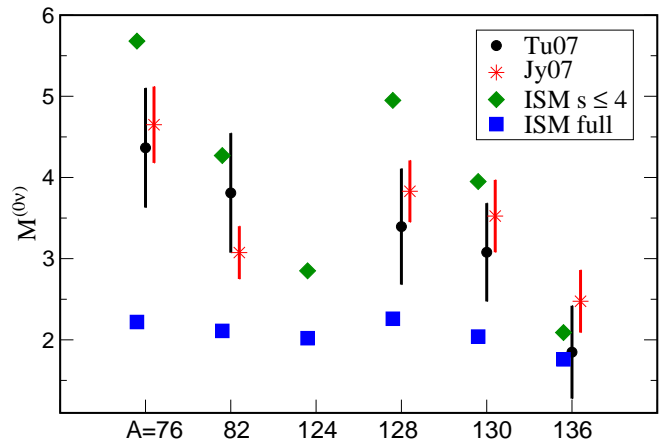}
\caption{\label{fig:m0nu}(color online) The neutrinoless double beta decay NME's; comparison of ISM and QRPA calculations.
 Tu07; QRPA results from ref.~\cite{rodin-err}. Jy07; QRPA results from ref.~\cite{kosu}. ISM s$\le$4 and ISM; present work.}
  \end{center}
\end{figure}

We can now proceed to compare in detail the ``state of the art'' ISM
and QRPA \cite{rodin-err,kosu} NME's in Fig.~\ref{fig:m0nu}. The QRPA results
for $^{124}$Sn are not yet available.  The range of QRPA values shown
in the figure is that given by the authors, and derives from the
different choices of g$_{pp}$ and g$_{A}$, as well as from their use or
not of a renormalized version of the QRPA. The larger values correspond to
g$_{A}$=1.25 and should therefore be preferred in the comparison with 
our predictions.  Both the QRPA and the ISM
calculations include the higher order corrections from
ref.~\cite{simko}.  For a proper comparison, the TU07 NME's have been
increased by 10\% due to their different choice of r$_0$. 
In all the calculations the short range correlations
 are modeled by the same Jastrow factor.

Several interesting conclusions stem from this figure. First, the fact
that the different QRPA calculations are now compatible. In addition,
they produce NME's that are strikingly close to the ISM ones
calculated at the truncation level s$\le$4.  In the
A=136 decay, in which the s$\le$4 truncation is a good approximation to
the full result, the QRPA values and the ISM ones do agree
(this seems to be also the case for the A=124 decay \cite{privat}).
  This suggests that, somehow, s$\le$4 is the implicit
truncation level of the QRPA.  In the QRPA calculations in a spherical
basis that we are discussing, the ground states of parent and grand
daughter, calculated in the BCS approximation, have generalized
seniority zero. The RPA ground state correlations of multipole
character (quadrupole, octupole, etc.), bring components with s$\ge$4
into these wave functions. But, for the RPA approximation to remain
valid, their amplitudes should decrease with s.  Indeed, in our ISM
s$\le$4 results, the percentage of s=0 components is always larger
than 70\%, a figure compatible with a QRPA description. However, in
the full calculation for the A=76, A=82, A=128, and A=130 decays, this
percentage can be as low as 25\% (actually, in  $^{76}$Se, the s=4
components almost double the percentage of the s=0 ones).
In these cases, the QRPA is bound
to overestimate the amount of s=0 components and, consequently, the
value of the NME's. In a sense, the QRPA can be said to be a ``low
seniority approximation'', roughly equivalent to the s$\le$4 ISM
truncations, that overestimate the NME's when the nuclei that
participate in the decay are strongly correlated by the multipole part
of the effective nuclear interaction.  The extent of the
overestimation depends on the degree of validity of the low seniority
approximation in each decaying pair.

 The values of M$^{(0\nu)}$ predicted by the present ISM calculation
for the  A=76, A=82, A=128, and A=130 decays, are smaller that the QRPA 
(central) ones by factors 1.5-2. Therefore, for a given value of the
effective neutrino mass, the predicted ISM half-lives of the 0$\nu\beta\beta$
decays are 2-4 times longer than the QRPA ones. Equivalently, for 
a given lower bound on the half-life, the ISM NME's produce upper
bounds on the effective neutrino mass that are larger than those of
the QRPA by factors 1.5-2. For instance, a bound  on 
T$_{\frac{1}{2}}$($^{76}$Ge $\rightarrow$ $^{76}$Se) of 10$^{25}$ y.
results in an effective neutrino mass of 1.05~eV with the ISM NME, and
500~meV with the QRPA one. The same bound for the half-life of the
 $^{130}$Te $\rightarrow$  $^{130}$Xe decay would lead to bounds on 
the neutrino mass of 410~meV and 270~meV respectively.

 In summary, we have analyzed the 0$\nu\beta\beta$ NME's in terms of
the J$^{\pi}$ of the decaying neutron pair.  We have found that in the
seniority zero limit the decays are strongly favored.  When the non
zero seniority components of the wave functions, originated by the
multipole terms of the nuclear effective interaction, are properly
taken into account, the matrix elements are drastically reduced.  In
particular, when the multipole correlations are large, the low seniority
truncations, s$\le$4, similar to those implicitly present in the
spherical QRPA approaches based in a BCS treatment of the pairing
interaction, are shown to overestimate the NME's.  Hence, we surmise
that, when the QRPA and ISM results do not agree,  the true NME's  should
be much closer to the ISM predictions than to the QRPA ones.


{\bf Acknowledgements} Supported by a grant of the Ministerio de  Educaci\'on y Ciencia (Spain),
                       FIS2006-01245, by the IN2P3(France)-CICyT(Spain) collaboration agreements,
                       and by the EU program  ILIAS N6 ENTApP WP1.

\end{document}